\documentclass[manuscript]{aastex}

\shorttitle{Heavy ion acceleration by SPA waves}
\shortauthors{Matsukiyo et al.}

\begin{document}


\title{Heavy ion acceleration by super Alfv\'{e}nic waves}


\author{S. Matsukiyo}
\affil{Faculty of Engineering Sciences, Kyushu University,\\
    Kasuga, Fukuoka, 816-8580, Japan}
\email{matsukiy@esst.kyushu-u.ac.jp}
\author{T. Akamizu}
\affil{Interdisciplinary Graduate School of Engineering Sciences, Kyushu University,\\
    Kasuga, Fukuoka, 816-8580, Japan}
\author{T. Hada}
\affil{Faculty of Engineering Sciences, Kyushu University,\\
    Kasuga, Fukuoka, 816-8580, Japan}
\email{hada@esst.kyushu-u.ac.jp}


\begin{abstract}
A generation mechanism of super Alfv\'{e}nic (SPA) waves in multi-ion species 
plasma is proposed and the associated heavy ion acceleration process is discussed. 
The SPA waves are thought to play important roles in 
particle acceleration since they have large wave electric field because 
of their high phase velocity. It is demonstrated by using full particle-in-cell simulation 
that large amplitude proton cyclotron waves, excited due to proton 
temperature anisotropy, nonlinearly destabilize SPA waves through parametric 
decay instability in a three component plasma composed of electrons, protons, 
and $\alpha$ particles. At the same time, $\alpha$ cyclotron waves get excited 
via another decay instability. A pre-accelerated $\alpha$ particle resonates 
simultaneously with the two daughter waves, the SPA waves and the $\alpha$ 
cyclotron waves, and it is further accelerated perpendicular to the ambient 
magnetic field. 
The process may work in astrophysical environments where sufficiently 
large temperature anisotropy of lower mass ions occurs.
\end{abstract}




\section{Introduction}

In many space and astrophysical environments observations show that heavy ions 
are preferentially accelerated. The ratio of galactic cosmic ray proton flux to helium 
flux decreases as rigidity increases \citep{aguilar15}. Increase of heavy ion 
abundance is commonly observed in solar energetic particle events (e.g., 
\cite{reames17,klecker07} and the references therein). In-situ observations 
in terrestrial magnetosphere  (e.g., \cite{kronberg14} and the references therein) 
and interplanetary space \citep{dayeh17,gruesbeck15,filwett17} also often detect 
preferentially accelerated heavy ions. 
While a number of theories 
have been proposed, the mechanism of heavy ion acceleration is still under 
debate. One of the extensively studied processes is resonant wave-particle 
interactions (cf., \cite{wang19,shevchenko06,tu03,gary03,horne97,thorne94,thorne93} 
and some other earlier works are reviewed in \cite{hollweg02}). 

\begin{figure}
\includegraphics[width=18pc]{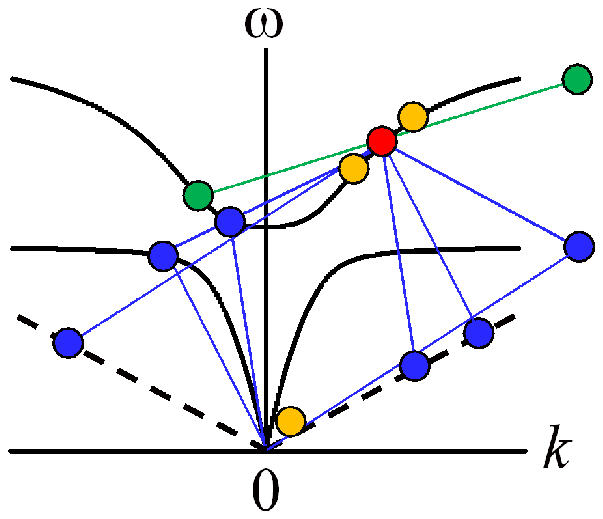}
\caption{Schematic $\omega-k$ diagram of a number of channels of parametric 
instabilities in a three species plasma.
\label{parametric}}
\end{figure}

About two decades ago, \cite{mizuta01} proposed a mechanism of preferential 
acceleration of $\alpha$ particles interacting with two left hand circularly 
polarized waves. They showed by performing a test particle simulation that an 
$\alpha$ particle is efficiently accelerated perpendicular to the ambient 
magnetic field if one of the two waves is on the branch of super Alfv\'{e}nic 
(SPA) waves in a three component plasma consisting of electrons, protons, 
and $\alpha$ particles. In the three component plasma there are two ion 
cyclotron wave branches in $\omega-k$ space as shown by the black solid 
lines in Fig.\ref{parametric}, i.e., proton cyclotron waves (upper branch) 
and $\alpha$ cyclotron waves (lower branch). 
The SPA waves are the high phase velocity, or small wavenumber, part of 
the proton cyclotron waves. Although the mechanism proposed by \cite{mizuta01} 
is efficient, this acceleration mechanism has not been paid much attention so far. 
It is probably because that a generation mechanism of the SPA waves has been 
unclear. 

In this study we propose a generation mechanism of the SPA waves 
through the nonlinear evolution of large amplitude lower phase velocity proton 
cyclotron waves. The latter waves are easily driven by proton temperature 
anisotropy known as the electromagnetic ion cyclotron (EMIC) instability. 
Here, we call this as proton EMIC instability. 
It is confirmed/expected that the anisotropy of proton temperature often 
becomes extremely high, $T_{p\perp} / T_{p\parallel} \gg 1$, 
in circumstances such as solar coronal hole \citep{isenberg19,markovskii07}, 
high Mach number perpendicular shock front \citep{shimada05,sckopke83}, 
intracluster medium \citep{santos14}, and so on. Here, $T_{p\perp}$ and
 $T_{p\parallel}$ denote proton temperature perpendicular and parallel to the 
ambient magnetic field, respectively. Linear growth rate of the proton EMIC 
instability in sufficiently large anisotropy is obtained as $\gamma \approx 
(\beta_{p\perp}/2)^{1/2}\Omega_p$, where $\beta_{p\perp}$ is perpendicular 
proton beta, the ratio of proton pressure perpendicular to the ambient magnetic 
field to magnetic pressure, and $\Omega_p$ denotes 
proton cyclotron frequency \citep{davidson75}. Therefore, if $\beta_{p\perp}$ 
becomes of the order of 0.1 or even higher, very strong proton EMIC 
instability is expected to be generated. 
In a nonlinear stage of such strong proton EMIC instability 
large amplitude proton EMIC 
waves can be the source of 
wave-wave interactions. By performing a full particle-in-cell (PIC) simulation, 
it is shown that the SPA waves are self-consistently generated in the 
late stage of the proton EMIC instability. We further demonstrate that 
$\alpha$ particles are preferentially accelerated in the simulation through 
the process discussed by \cite{mizuta01}. 

%

\section{Nonlinear generation of super Alfv\'{e}nic waves}

We propose a parametric instability of a proton EMIC wave as a mechanism 
of SPA wave generation. If once large amplitude proton EMIC 
waves get excited, they can be a source of parametric instabilities. 
The lowest order parametric instability occurs through three wave interactions 
in which the resonance conditions, $\omega_3 = \omega_1 \pm \omega_2, 
k_3 = k_1 \pm k_2$, are fulfilled among the interacting waves. Here, $\omega_{1,2,3}$ 
and $k_{1,2,3}$ are the frequencies and wavenumbers of the waves.
In a three 
component electron-proton-$\alpha$ plasma many channels of parametric 
instabilities exist. Fig.\ref{parametric} schematically indicates possible channels 
of parametric 
instabilities in $\omega-k$ diagram, when a parent proton EMIC wave 
(denoted by a red circle) and all daughter waves propagate along the ambient 
magnetic field 
(Note that if the parent wave is on the same branch with negative $k$, all 
daughter waves have wavenumbers with opposite signs.).
In addition to the two transverse wave branches (proton and 
$\alpha$ cyclotron waves), a longitudinal wave branch (ion acoustic waves) 
is denoted by the dashed line. For example, the yellow circles are the 
possible daughter waves generated due to a modulational instability. The green 
circles are another possible daughter waves due to a beat instability. In addition, 
a number of decay instabilities are possible to be generated. In the figure three 
channels of decay instability are represented by the blue parallelograms. 
(The above mentioned resonance conditions lead to the parallelograms in the 
$\omega-k$ space.)
One of 
them has a daughter wave on the proton cyclotron branch. This daughter 
wave has rather small wavenumber and its phase velocity exceeds the 
Alfv\'{e}n velocity, i.e., SPA wave. 
This channel of decay instability was found by \cite{gomberoff95} in their 
linear dispersion analysis. 
We will see this channel of decay instability is generated 
in the nonlinear stage of proton EMIC instability in the PIC simulation below.

\section{Simulation}

A standard periodic one-dimensional PIC simulation of three species plasma is performed. 
The system size is $L/(c/\omega_{pp})=327.68$, where $c$ is the speed of light and 
$\omega_{pp}$ is the proton plasma frequency in the case of no $\alpha$ particles. The 
size of spatial grid is the electron Debye length, while 
the number of super particles per cell for each species is 200. The 
ambient magnetic field is along the $x$-axis. The ratio of electron cyclotron frequency 
to plasma frequency is $\Omega_e / \omega_{pe} = 0.5$. The mass ratio of the three 
species is $m_e : m_p : m_{\alpha} = 1 : 25 : 100$, where $m_e$, $m_p$, and $m_{\alpha}$ 
are the mass of electrons, protons, and $\alpha$ particles, 
respectively. The relative number density of the $\alpha$ particles to the electrons 
is $n_{\alpha} / n_e = 0.1$. 
Perpendicular proton beta is $\beta_{p\perp} = 0.2$ and proton temperature 
anisotropy is $T_{p\perp} / T_{p\parallel}=100$. The temperature of $\alpha$ 
particles is isotropic and the same as that of parallel proton temperature, 
$T_{\alpha} = T_{p\parallel}$. While isotropic electron beta is $\beta_e = 0.08$, 
we confirmed that the following results are almost independent of electron beta. 
The proton temperature anisotropy is the only source 
of free energy in this system so that proton EMIC waves are the only waves that 
are linearly unstable.

\begin{figure}
\includegraphics[width=18pc]{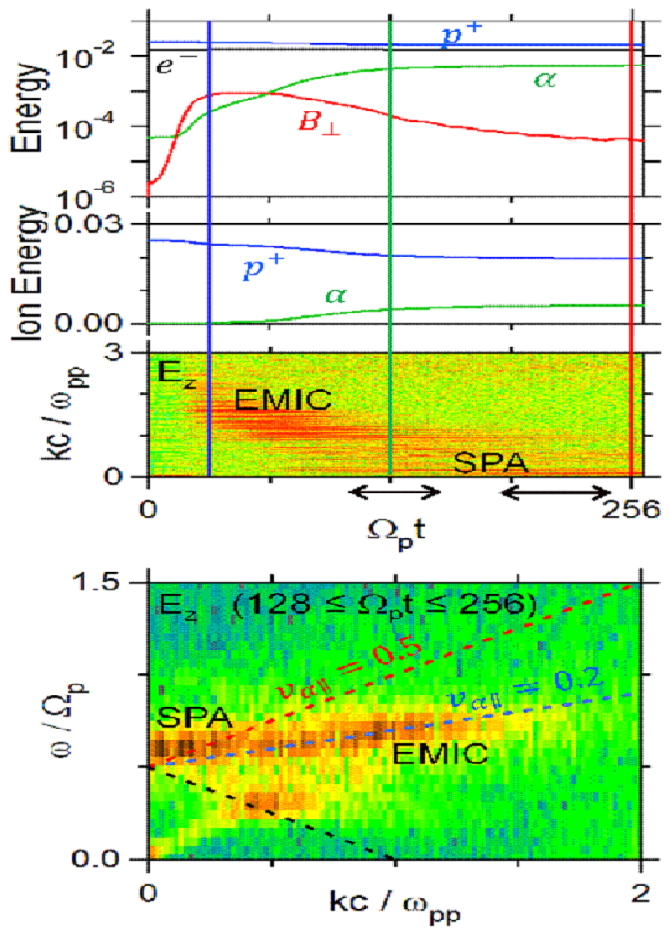}
\caption{Time history of PIC simulation. The top panel denotes energy time history of 
($p^+$) protons, ($e^-$) electrons, ($\alpha$) $\alpha$ particles, and ($B_{\perp}$) 
transverse magnetic field, respectively, in the logarithmic 
scale. The second panel shows only the energy of protons and $\alpha$ particles in 
the linear scale. The third panel indicates the time evolution of wavenumber spectrum 
of $E_z$ component. The $\omega-k$ spectrum of $E_z$ in the time interval of 
$128 \le \Omega_p t \le 256$ is shown in the fourth panel.
\label{pic_thistory}}
\end{figure}
Fig.\ref{pic_thistory} represents time evolution of energy (the first and the second panels) 
and wavenumber spectrum of $E_z$ component (the third panel). The time is normalized 
to the inverse proton cyclotron frequency, $\Omega^{-1}_p$. The rapid increase of the 
transverse magnetic field ($B_{\perp}$) energy is due to the proton EMIC instability. 
This leads 
to relatively broad band spectrum of $E_z$ component as well as the small increase 
of the $\alpha$ particle energy until that the transverse magnetic field energy is 
saturated. After that, the energy of the $\alpha$ particles increases more and the 
spectral peak of the proton EMIC waves shifts to smaller wavenumber. Furthermore, an 
additional spectral peak appears in very small wavenumber, $kc/\omega_{pp} \ll 1$. 
The amplified waves are the SPA waves. This is confirmed in the fourth panel in which 
$\omega-k$ spectrum of the left-hand polarized fluctuations 
($\omega>0$) 
of $E_z$ component 
corresponding to the time interval of $128 \le \Omega_p t \le 256$ is depicted. 
Here, we have used the technique of Fourier decomposition to extract only the 
fluctuations with positive helicity ($k>0$) and frequency ($\omega>0$) from 
the original $E_z$ data after \cite{terasawa86}.
Electron energy is almost unchanged 
throughout the run indicating that they do not contribute to the process discussed 
above.

\begin{figure}
\includegraphics[width=20pc]{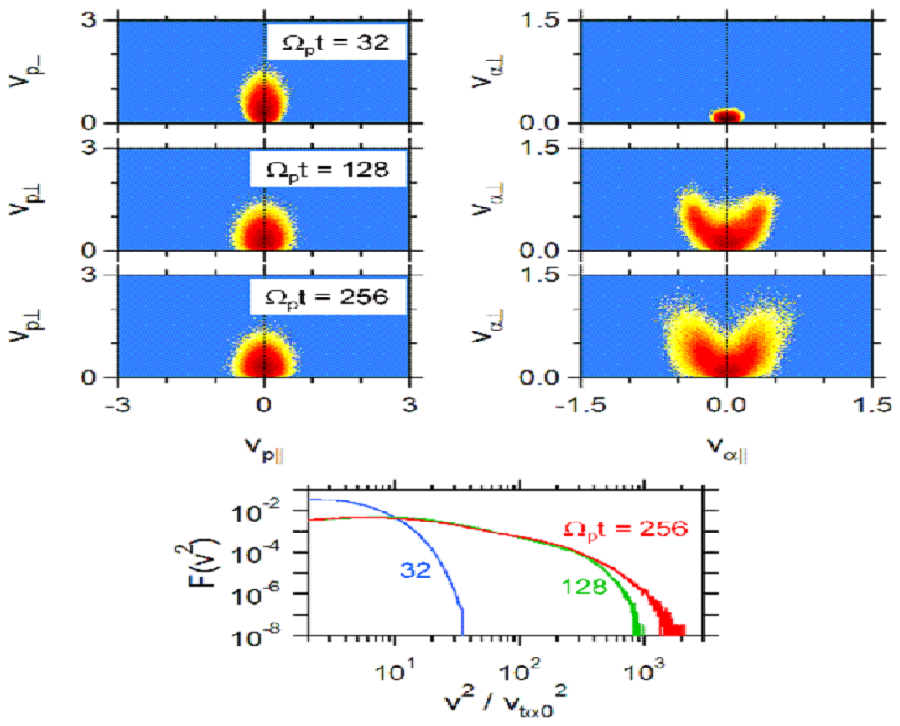}
\caption{Evolution of velocity and energy distribution functions. The top three panels 
show velocity distribution functions of protons (left) and $\alpha$ particles (right) at 
$\Omega_p t = 32, 128$ and $256$, respectively. The bottom panel denotes energy 
distribution functions of $\alpha$ particles at the corresponding times. The energy 
is normalized to initial thermal energy of the $\alpha$ particles.
\label{pic_distfunc}}
\end{figure}

In Fig.\ref{pic_distfunc} the top three panels show velocity distribution functions of 
protons (left) and $\alpha$ particles (right) at three different times, $\Omega_p t = 32, 
128$ and $256$, indicated by the vertical lines in Fig.\ref{pic_thistory}. The velocities parallel 
and perpendicular to the ambient magnetic field is normalized to Alfv\'{e}n velocity, $v_A$, 
in the case of no $\alpha$ particles. As time passes, the protons are 
isotropized and the $\alpha$ particles are accelerated to form V-shaped distribution 
in the velocity space. \cite{tanaka85} discussed this V-shaped velocity distribution of the 
$\alpha$ particles by using a hybrid simulation. He used the term 'heating' instead of 
'acceleration' to express the V-shaped distribution. However, it is obvious from the 
bottom panel that the energy distribution function of the $\alpha$ particles indicates 
nonthermal feature when the V-shaped velocity distribution is seen. \cite{tanaka85} 
further interpreted 
this heating as a result of nonlinear interactions between the $\alpha$ particles and the 
proton EMIC waves. 


\begin{figure}
\includegraphics[width=18pc]{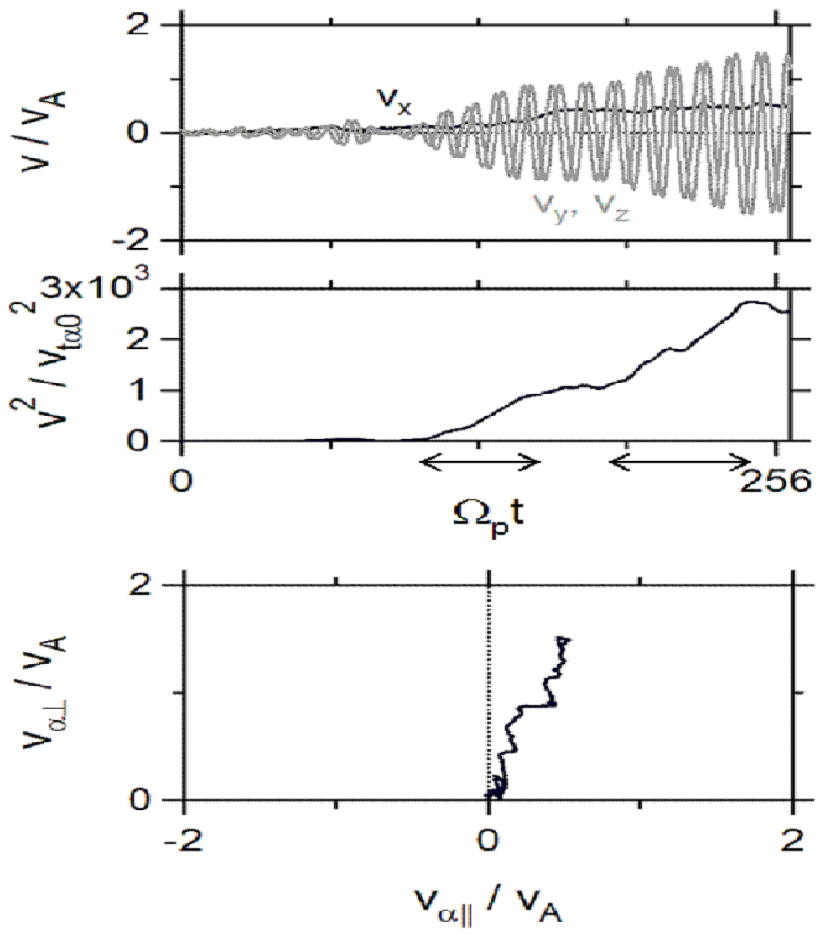}
\caption{Trajectory of a nonthermal $\alpha$ particle. From the top, time evolution 
of three velocity components, that of energy, and the trajectory in velocity space 
are plotted. There are two main acceleration phases indicated by the horizontal 
arrows in the middle panel. The energy 
is normalized to initial thermal energy of the $\alpha$ particles.
\label{pic_acc}}
\end{figure}

Here, we focus on a trajectory of one of the well accelerated nonthermal $\alpha$ 
particles. In Fig.\ref{pic_acc} the top and the middle panels show the time evolution 
of three components of velocity and that of energy. In the bottom panel a 
corresponding trajectory in velocity space is depicted. It is recognized in the 
middle panel that there are two acceleration phases. The first acceleration phase 
occurs in $105 \le \Omega_p t \le 155$, while the second phase occurs in 
$185 \le \Omega_p t \le 245$. The corresponding time domains are indicated by 
horizontal arrows in both Fig.\ref{pic_acc} and Fig.\ref{pic_thistory}. In the both 
acceleration phases this $\alpha$ particle gains energy mainly perpendicular to 
the ambient magnetic field as one can read from the top panel of Fig.\ref{pic_acc}. 
In the first acceleration phase its parallel velocity is about $0.2v_A$ in average. 
The blue dashed line in the bottom panel of Fig.\ref{pic_thistory} denotes the cyclotron 
resonance condition of the $\alpha$ particle 
($\omega - k v_{\alpha \parallel} = \Omega_{\alpha}$) with its parallel velocity of 
$v_{\alpha \parallel} = 0.2 v_A$. The line overlaps the region of spectral peak amplified 
by the proton EMIC instability. Therefore, it is natural to conclude that the acceleration 
in the first phase is due to the interaction with the proton EMIC waves. After this first 
phase, the parallel velocity increases roughly to $0.5 v_A$ probably through pitch angle 
scattering while its energy is not so changed. In the second acceleration phase the 
particle energy increases while this parallel velocity is almost kept constant. The 
corresponding resonance condition with $v_{\alpha \parallel} = 0.5 v_A$ is also 
shown as the red dashed line in the bottom panel of Fig.\ref{pic_thistory}. It is indicated 
that this particle can resonate with the SPA waves. Since the wave spectrum is 
essentially symmetric with respect to $k$, the red line is folded back and indicated by 
the black dashed line. On this dashed line, there is another spectral peak around 
$kc / \omega_{pp} = 0.5$, which is due to another decay instability (Fig.\ref{parametric}) 
on the $\alpha$ cyclotron branch. Hence, in this particular case the particle can 
resonate with the two kinds of nonlinearly generated waves, the SPA waves and the 
$\alpha$ cyclotron waves, which is the same situation as discussed by 
\cite{mizuta01}.

\section{Summary and Discussions}

We proposed a mechanism of nonlinear generation of SPA waves in a 
three component plasma, which is based on a parametric instability of 
proton cyclotron waves generated by proton EMIC instability. The process
was demonstrated self-consistently in a one-dimensional full PIC simulation. 
In the simulation not only the SPA waves but also the $\alpha$ cyclotron 
waves get excited. While protons lose their initial free energy and are 
isotropized, some $\alpha$ particles are accelerated to nonthermal energy. 
During its acceleration process, 
an $\alpha$ particle resonates simultaneously with 
the nonlinearly generated SPA waves and the $\alpha$ cyclotron waves. 
This is essentially the same situation as discussed by \cite{mizuta01}.  


The generation of SPA waves in the nonlinear stage of the proton EMIC 
instability occurs in a wide parameter range. We confirmed that they are 
indeed generated even for smaller proton temperature anisotropy 
($T_{p\perp} / T_{p\parallel} = 20$), smaller frequency ratio 
($\Omega_e / \omega_{pe} = 0.1$), and for larger mass ratio 
($m_e:m_p:m_{\alpha}=1:100:400$) cases.

In the current simulation $\alpha$ particles are energized first by interacting 
with the proton EMIC waves 
which have relatively lower phase velocities. Later, some of the energized 
particles resonate with both the nonlinearly generated SPA waves and 
the $\alpha$ cyclotron waves. Hence, the proton EMIC waves play also 
a role in injection, i.e., pre-acceleration of the $\alpha$ particles to be further 
able to resonate with the SPA as well as the $\alpha$ cyclotron waves. 
There may be other injection mechanisms. In this study initial free energy 
of the system is provided only by the proton temperature anisotropy to 
make the scenario proposed here more visible. However, a situation that 
both protons and $\alpha$ particles have temperature anisotropy is also 
plausible. Since the temperature anisotropy of $\alpha$ particles results 
in generation of $\alpha$ EMIC instability, some $\alpha$ particles can 
be resonantly accelerated by self-generated $\alpha$ EMIC waves so that 
injected into further interaction with the SPA waves. These $\alpha$ EMIC 
waves may also play an alternative role of the $\alpha$ cyclotron waves 
generated through a decay instability of proton EMIC waves. 

Finally, we discuss the effect of relative $\alpha$ particle density, 
$n_{\alpha}/n_e$, which may affect the dispersion property of the interacting 
waves as well as the acceleration of $\alpha$ particles. When $n_{\alpha}/n_e$ 
increases, proton density decreases so that 
free energy of the proton EMIC instability decreases even if proton temperature 
anisotropy is unchanged. This may suppress the acceleration of $\alpha$ particles. 
On the other hand, $n_{\alpha}/n_e$ influences the topology of $\omega-k$ 
diagram. For instance, the cutoff frequency of the proton cyclotron branch 
increases with $n_{\alpha}/n_e$. In such a case it is inferred from 
Fig.\ref{parametric} that the generated SPA wave has higher phase velocity 
or equivalently stronger wave electric field. This may contribute to efficient 
acceleration of $\alpha$ particles. Hence, there may be the value of 
$n_{\alpha}/n_e$ leading to highest efficiency of acceleration. To confirm this 
will be the future issue.

\acknowledgments


We thank M. Hoshino, T. Amano, and Y. Matsumoto for fruitful discussions.



\end{document}